\documentclass[a4paper,10pt]{article}

\usepackage{bm,amsmath,amssymb,graphicx,mathrsfs}

\newcommand{\nc}{\newcommand}
\nc{\rnc}{\renewcommand}
\nc{\nn}{\nonumber}
\nc{\bra}{\langle}
\nc{\ket}{\rangle}
\rnc{\[}{\left[}
\rnc{\]}{\right]}

\nc{\sh}{\mathrm{sh}}
\nc{\ch}{\mathrm{ch}}

\begin{document}

\title{Boundary correlation functions of the six and
nineteen vertex models with domain wall boundary conditions}

\author{Kohei Motegi \\
Okayama Institute for Quantum Physics, \\
Kyoyama 1-9-1,
Okayama 700-0015, Japan}

\maketitle

\begin{abstract}
Boundary correlation functions of the 
six and nineteen vertex models
on an $N \times N$ lattice with
domain wall boundary conditions are studied.
The general expression of the boundary correlation functions
is obtained for the six vertex model by use of the quantum inverse
scattering method.
For the nineteen vertex model,
the boundary correlation functions are shown
to be expressed in terms of
those for the six vertex model.
\\ 
\\
Keywords: integrable vertex model,
quantum inverse scattering method, correlation function
\end{abstract}

\section{Introduction}
The six vertex model is 
one of the most fundamental exactly solved models
in statistical physics \cite{Slater,Lieb,Sutherland,Baxter}.
Not only the periodic boundary condition
but also the domain wall boundary condition
is an interesting boundary condition.
For example, the partition function is deeply 
related to the norm \cite{Korepin} and the scalar product
\cite{Slavnov} of the XXZ chain.
The determinant formula of the partition function \cite{Izergin,ICK}
lead Slavnov \cite{Slavnov} to obtain a compact representation of
the scalar product, which plays a fundamental role
in calculating correlation functions of the XXZ chain
\cite{KIEU,KIB,KMT,KMST}.
The determinant formula also led to a deep advance
in enumerative combinatorics
\cite{Zeilberger,Kuperberg,Bressoud}.
For example, it was used to give a concise proof of the 
numbers of the alternating sign matrices
for a given size.
Recently, the correspondences between the partition function
and the Schur polynomial \cite{La} and KP $\tau$ function
\cite{FWZ} have been revealed.
The determinant representations of partition functions have been
extended to other models such as the higher spin vertex models
\cite{CFK}, 
Felderhof models \cite{CFWZ} and so on.
The domain wall boundary conditions are also interesting from
the physical point of view since it exhibits phase separation phenomena
\cite{SZ,AR,CPphase,CPphase2,Reshetikhin}.

The calculation of correlation functions are also
interesting in the domain wall boundary condition itself.
Several kinds of them such as the boundary one point functions,
two point functions, boundary polarization
\cite{BKZ,BPZ,FP,CP}
and the emptiness formation probability \cite{CP2} have been calculated.

In this paper, we calculate
boundary correlation functions for the six and nineteen vertex models
on an $N \times N$ lattice with domain wall boundary condition.
For the six vertex model,
we use the quantum inverse scattering method,
and apply the approach of \cite{CP2} to build
and solve two recursive relations for the boundary
correlation functions, providing for them a general expression.
The boundary correlation functions we consider
includes the boundary polarization and
boundary emptiness formation probability (EFP) as
special cases.

Next, by use of fusion, we show that the boundary correlation functions
for the nineteen vertex (Fateev-Zamolodchikov \cite{FZ})
model can be reduced to those for the six vertex model.
In particular, the EFP of length $s$ for the nineteen vertex model reduces to 
that of length $2s$ for the six vertex model.

The outline of this paper is as follows.
In the next section, we define the six vertex model
with domain wall boundary condition.
The general expression for the boundary correlation functions
is obtained in section 3. In section 4, the boundary correlation
functions of the nineteen vertex model are considered by use of
of fusion. The emptiness formation probability for the
nineteen vertex model is expressed in the determinant form
in the homogeneous limit in section 5.

\section{Six vertex model}
The six vertex model is a model in statistical mechanics,
whose local states
are associated with edges of a square lattice,
which can take two values.
The Boltzmann weights are assigned to its vertices,
and each weight is determined by the configuration
around a vertex. What plays the fundamental role is the $R$-matrix
\begin{align}
R(\lambda, \nu)
=&\left(
\begin{array}{cccc}
1 & 0 & 0 & 0 \\
0 & \frac{ \sh (\lambda-\nu)}{ \sh (\lambda-\nu+\eta)} & 
\frac{ \sh \eta}{ \sh (\lambda-\nu+\eta)} & 0 \\
0 & \frac{ \sh \eta}{ \sh (\lambda-\nu+\eta)} &
\frac{ \sh (\lambda-\nu)}{ \sh (\lambda-\nu+\eta)} & 0 \\
0 & 0 & 0 & 1
\end{array}
\right),
\end{align}
which satisfies the Yang-Baxter equation
\begin{align}
R_{12}(\lambda, \nu) R_{13}(\lambda, \mu) R_{23}(\nu, \mu)
=R_{23}(\nu, \mu) R_{13}(\lambda, \mu) R_{12}(\lambda, \nu).
\label{YangBaxter}
\end{align}
We consider the six vertex model
on a $N \times N$ lattice depicted in Figure.
The spins are aligned all up at the bottom and right boundaries,
and all down at the top and left boundaries.
At the intersection of the $\alpha$-th row (from the bottom)
and the $k$-th column (from the left), we associate the statistical weight
\begin{align}
\mathcal{L}_{\alpha k}(\lambda_\alpha,
\nu_k)&=\sh(\lambda_{\alpha}-\nu_k+\eta/2)
R_{\alpha k}(\lambda_{\alpha}-\eta/2, \nu_k) \nn \\
&=\left(
\begin{array}{cccc}
a(\lambda_{\alpha}, \nu_k) & 0 & 0 & 0 \\
0 & b(\lambda_{\alpha}, \nu_k) & c & 0 \\
0 & c & b(\lambda_{\alpha}, \nu_k) & 0 \\
0 & 0 & 0 & a(\lambda_{\alpha}, \nu_k)
\end{array}
\right),
\end{align}
where
\begin{align}
a(\lambda, \nu)=\sh(\lambda-\nu+\eta/2), 
\ b(\lambda, \nu)=\sh (\lambda-\nu-\eta/2),
\ c=\sh \eta.
\end{align}
We refer to the $\alpha$-th row as the auxiliary space 
$\mathcal{V}_\alpha$ and the $k$-th column as the quantum space
$\mathcal{H}_k$. Let us denote 
$\{ \lambda \}=\{\lambda_1, \lambda_2, \dots , \lambda_N  \},
\{ \nu \}=\{\nu_1, \nu_2, \dots , \nu_N  \}$,
and the basis (dual basis) of the spin-1/2 representation
as $|+ \rangle, |- \rangle$ ($\langle +|, \langle -|$). 

The partition function of the six vertex model, which is the summation
of products of statistical weights over all possible configurations
can be formally represented as
\begin{align}
\mathcal{Z}_N(\{ \lambda \}, \{ \nu \})
= {}_a\langle +|| {}_q\langle -|| \prod_{\alpha,k=1}^N \mathcal{L}_{\alpha k}
(\lambda_{\alpha}, \nu_k)||- \rangle_a ||+ \rangle_q,
\end{align}
where
$ ||+ \rangle=\otimes_{k=1}^N | + \rangle_k ,
  ||- \rangle=\otimes_{k=1}^N | - \rangle_k,
\langle + ||=\otimes_{k=1}^N  {}_k \langle +|,
\langle - ||=\otimes_{k=1}^N  {}_k \langle -|, $
and we distinguish the spins on the quantum and
auxiliary spaces by the subscripts "$q$" and "$a$".
\begin{figure}[htbp]
 \begin{center}
  \includegraphics[width=100mm]{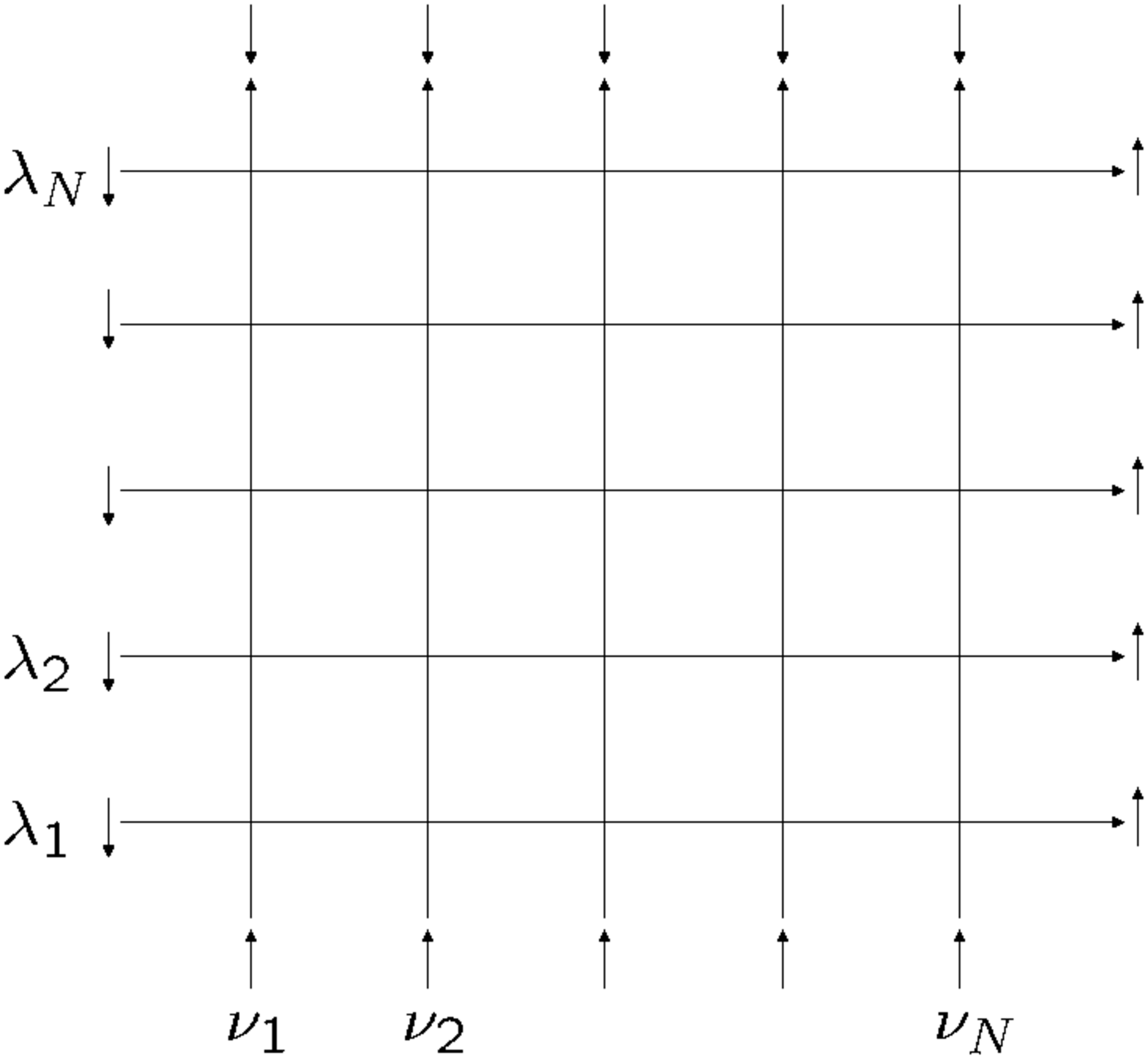}
 \end{center}
 \caption{The six vertex model with domain wall boundary condition.}
 \label{fig:one}
\end{figure}
The partition function has the following determinant form
\cite{Izergin,ICK}
\begin{align}
\mathcal{Z}_{N}(\{ \lambda \}, \{ \nu \})
=\frac{
\prod_{\alpha=1}^N \prod_{k=1}^N 
a(\lambda_\alpha, \nu_k)b(\lambda_\alpha, \nu_k)
\det M(\{ \lambda \}, \{ \nu \})
}
{
\prod_{1 \le \alpha < \beta \le N}d(\lambda_\beta, \lambda_\alpha)
\prod_{1 \le j < k \le N} d(\nu_j, \nu_k)
},
\label{partition}
\end{align}
where
\begin{align}
d(\lambda, \nu)=\sh(\lambda-\nu), \
M_{\alpha k}=\varphi(\lambda_\alpha, \nu_k), \
\varphi(\lambda, \nu)
=\frac{c}{a(\lambda, \nu)b(\lambda, \nu)}.
\end{align}
Introducing the monodromy matrix
\begin{align}
T_\alpha(\lambda_{\alpha}, \{ \nu \})=&\mathcal{L}_{\alpha N}
(\lambda_{\alpha},\nu_N)
\cdots \mathcal{L}_{\alpha 1}(\lambda_{\alpha},\nu_1) \nonumber \\
=&\left(
\begin{array}{cc}
A(\lambda_{\alpha}, \{ \nu \})  & B(\lambda_{\alpha}, \{ \nu \}) \\
C(\lambda_{\alpha}, \{ \nu \}) & D(\lambda_{\alpha}, \{ \nu \})
\end{array}
\right), \label{onerow}
\end{align}
the partition function can be represented as
\begin{align}
\mathcal{Z}_{N}(\{ \lambda \}, \{ \nu \})
={}_q \langle -|| B(\lambda_{N}, \{ \nu \}) \cdots
B(\lambda_{1}, \{ \nu \})  ||+ \rangle_q \,
\end{align}
From the Yang-Baxter equation, one has
\begin{align}
R_{\alpha \beta}(\mu-\lambda)T_\alpha(\mu, \{ \nu \})
T_\beta(\lambda, \{ \nu \})
=T_\beta(\lambda, \{ \nu \})T_\alpha(\mu, \{ \nu \})
R_{\alpha \beta}(\mu-\lambda).
\label{RTT}
\end{align}
From \eqref{RTT}, one has
\begin{align}
A(\lambda, \{ \nu \})B(\mu, \{ \nu \})
&=f(\lambda, \mu)B(\mu, \{ \nu \})A(\lambda, \{ \nu \})
+g(\mu, \lambda)B(\lambda, \{ \nu \})A(\mu, \{ \nu \}), \\
B(\lambda, \{ \nu \})A(\mu, \{ \nu \})
&=f(\lambda, \mu)A(\mu, \{ \nu \})B(\lambda, \{ \nu \})
+g(\mu, \lambda)A(\lambda, \{ \nu \})B(\mu, \{ \nu \}), \\
B(\lambda, \{ \nu \})B(\mu, \{ \nu \})
&=B(\mu, \{ \nu \})B(\lambda, \{ \nu \}),
\label{RTTrelations}
\end{align}
where
\begin{align}
f(\mu, \lambda)=\frac{\sh(\lambda-\mu+\eta)}{\sh(\lambda-\mu)},
\
g(\mu, \lambda)=\frac{\sh \eta}{\sh(\lambda-\mu)},
\end{align}
for example.
\section{Boundary correlation functions of the six vertex model}
In this section, we consider the following
boundary correlation functions
\begin{align}
\mathcal{F}_N^{(r, \epsilon_1, \cdots, \epsilon_{s})}
(\{ \lambda \}, \{ \nu \})
&=\frac{
\widetilde{\mathcal{F}}_N^{(r, \epsilon_1, \cdots, \epsilon_{s})}
(\{ \lambda \}, \{ \nu \})
}
{\mathcal{Z}_{N}(\{ \lambda \}, \{ \nu \})},
\label{boundarycorrelation}
\\
\widetilde{\mathcal{F}}_N^{(r, \epsilon_1, \cdots, \epsilon_{s})}
(\{ \lambda \}, \{ \nu \})
&= {}_a\langle +|| {}_q\langle -|| 
\prod_{\alpha=r+1}^N \prod_{k=1}^N \mathcal{L}_{\alpha k}
(\lambda_{\alpha}, \nu_k)
\prod_{k=1}^s \pi_k^{\epsilon_k}
\prod_{\alpha=1}^r \prod_{k=1}^N \mathcal{L}_{\alpha k}
(\lambda_{\alpha}, \nu_k)
||- \rangle_a ||+ \rangle_q,
\label{numerator}
\end{align}
where $\pi_k^+=|+ \rangle_k {}_k \langle +|$ and $
\pi_k^-=|- \rangle_k {}_k \langle -|$ is a projection onto the
up and down spin respectively.
Some special cases of 
this general boundary correlation function reduces to the ones previously
considered \cite{BKZ,BPZ,FP,CP,CP2}.
We calculate the boundary correlation functions
by use of the quantum inverse scattering method,
applying the approach of \cite{CP2}.
\begin{figure}[htbp]
 \begin{center}
  \includegraphics[width=100mm]{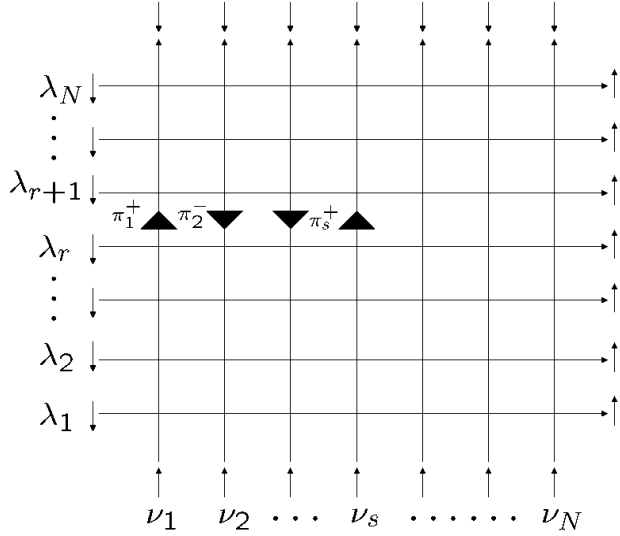}
 \end{center}
 \caption{An example of boundary correlation function.}
\end{figure}

First,
note that \eqref{numerator} can be 
expressed as
\begin{align}
\widetilde{\mathcal{F}}_N^{(r, \epsilon_1, \cdots, \epsilon_{s})}
(\{ \lambda \}, \{ \nu \})
= {}_q\langle -|| 
B(\lambda_{N}, \{ \nu \}) \cdots B(\lambda_{r+1}, \{ \nu \})
\prod_{k=1}^s \pi_k^{\epsilon_k}
B(\lambda_{r}, \{ \nu \}) \cdots B(\lambda_{1}, \{ \nu \})
||+ \rangle_q,
\end{align}
in the quantum inverse scattering language.

We introduce the following two-site model \cite{KIB}
in order to obtain recursive relations between boundary
correlation functions of different lattice sizes.
\begin{align}
T(\lambda, \{ \nu \})=&T_2(\lambda, \{ \nu \} \backslash \nu_1)
T_1(\lambda, \nu_1) \\
T_2(\lambda, \{ \nu \} \backslash \nu_1)=&\mathcal{L}_{\alpha N}
(\lambda,\nu_N)
\cdots \mathcal{L}_{\alpha 2}(\lambda,\nu_2) \nonumber \\
=&\left(
\begin{array}{cc}
A_2(\lambda, \{ \nu \} \backslash \nu_1)  & 
B_2(\lambda, \{ \nu \} \backslash \nu_1) \\
C_2(\lambda, \{ \nu \} \backslash \nu_1) &
D_2(\lambda, \{ \nu \} \backslash \nu_1)
\end{array}
\right), \\
T_1(\lambda, \nu_1)=&\mathcal{L}_{\alpha 1}
(\lambda, \nu_1).
\end{align}
Applying
\begin{align}
{}_1 \langle +|B(\lambda, \{ \nu \})|+ \rangle_1
&=b(\lambda, \nu_1)B_2(\lambda, \{ \nu \} \backslash \nu_1), \nonumber \\
{}_1 \langle -|B(\lambda, \{ \nu \})|+ \rangle_1
&=cA_2(\lambda, \{ \nu \} \backslash \nu_1), \nonumber \\
{}_1 \langle +|B(\lambda, \{ \nu \})|- \rangle_1
&=0, \nonumber \\
{}_1 \langle -|B(\lambda, \{ \nu \})|- \rangle_1
&=a(\lambda, \nu_1) B_2(\lambda, \{ \nu \} \backslash \nu_1),
\end{align}
iteratively, one has
\begin{align}
&{}_1 \langle +| B(\lambda_n, \{ \nu \}) \cdots B(\lambda_1, \{ \nu \})
|+ \rangle _1=\prod_{j=1}^n b(\lambda_j, \nu_1)
B_2(\lambda_n, \{ \nu \} \backslash \nu_1) \cdots 
B_2(\lambda_1, \{ \nu \} \backslash \nu_1), \label{rel} \\
&{}_1 \langle -| B(\lambda_n, \{ \nu \}) \cdots B(\lambda_1, \{ \nu \})
|- \rangle _1=\prod_{j=1}^n a(\lambda_j, \nu_1)
B_2(\lambda_n, \{ \nu \} \backslash \nu_1) \cdots 
B_2(\lambda_1, \{ \nu \} \backslash \nu_1), \label{rel2} \\
&{}_1 \langle -| B(\lambda_n, \{ \nu \}) \cdots B(\lambda_1, \{ \nu \})
|+ \rangle _1  \nonumber \\
=&
\sum_{\alpha=1}^n \prod_{\beta=\alpha+1}^n a(\lambda_\beta, \nu_1)
c \prod_{\beta=1}^{\alpha-1}b(\lambda_\beta, \nu_1) 
B_2(\lambda_n, \{ \nu \} \backslash \nu_1) \cdots 
B_2(\lambda_{\alpha+1}, \{ \nu \} \backslash \nu_1) \nonumber \\
&\times 
A_2(\lambda_\alpha, \{ \nu \} \backslash \nu_1)
B_2(\lambda_{\alpha-1}, \{ \nu \} \backslash \nu_1) \cdots 
B_2(\lambda_1, \{ \nu \} \backslash \nu_1).
\label{relation}
\end{align}
Combining \eqref{relation},
\begin{align}
A_2(\lambda, \{ \nu \} \backslash \nu_1)
B_2(\mu, \{ \nu \} \backslash \nu_1)
=&f(\lambda, \mu)
B_2(\mu, \{ \nu \} \backslash \nu_1)
A_2(\lambda, \{ \nu \} \backslash \nu_1) \nonumber \\
&+g(\mu, \lambda)
B_2(\lambda, \{ \nu \} \backslash \nu_1)
A_2(\mu, \{ \nu \} \backslash \nu_1), 
\end{align}
and
\begin{align}
A_2(\lambda, \{ \nu \} \backslash \nu_1) \otimes_{k=2}^N|+ \rangle_k
=\prod_{k=2}^N a(\lambda, \nu_k) \otimes_{k=2}^N|+ \rangle_k,
\end{align}
we get
\begin{align}
&{}_1 \langle -| B(\lambda_r, \{ \nu \}) \cdots B(\lambda_1, \{ \nu \})
||+ \rangle \nonumber \\
=&
\sum_{\alpha=1}^r c
\prod_{ \substack{
\beta=1 \\
\beta \neq \alpha }}^r b(\lambda_\beta, \nu_1)
\prod_{ \substack{
\beta=1 \\
\beta \neq \alpha}}^{r} f(\lambda_\alpha, \lambda_\beta)
\prod_{k=2}^N a(\lambda_\alpha, \nu_k)
\prod_{ \substack{k=1 \\ k \neq \alpha}}^r
B_2(\lambda_k, \{ \nu \} \backslash \nu_1)
\otimes_{k=2}^N|+ \rangle_k.
\label{relation2}
\end{align}
In the same way as \eqref{relation2},
we can also show the following relation
\begin{align}
& \langle -|| B(\lambda_N, \{ \nu \}) \cdots B(\lambda_{r+1}, \{ \nu \})
|+ \rangle_1 \nonumber \\
=&
\sum_{\alpha=r+1}^N c
\prod_{ \substack{
\beta=r+1 \\
\beta \neq \alpha }}^N a(\lambda_\beta, \nu_1)
\prod_{ \substack{
\beta=r+1 \\
\beta \neq \alpha}}^{N} f(\lambda_\beta, \lambda_\alpha)
\prod_{k=2}^N b(\lambda_\alpha, \nu_k)
\otimes_{k=2}^N {}_k \langle -|
\prod_{ \substack{k=r+1 \\ k \neq \alpha}}^N
B_2(\lambda_k, \{ \nu \} \backslash \nu_1),
\label{relation3}
\end{align}
utilizing
\eqref{relation},
\begin{align}
B_2(\lambda, \{ \nu \} \backslash \nu_1)
A_2(\mu, \{ \nu \} \backslash \nu_1)
=&f(\lambda, \mu)
A_2(\mu, \{ \nu \} \backslash \nu_1)
B_2(\lambda, \{ \nu \} \backslash \nu_1) \nonumber \\
&+g(\mu, \lambda)
A_2(\lambda, \{ \nu \} \backslash \nu_1)
B_2(\mu, \{ \nu \} \backslash \nu_1)
, 
\end{align}
and
\begin{align}
\otimes_{k=2}^N  {}_k \langle -|
A_2(\lambda, \{ \nu \} \backslash \nu_1) 
=\prod_{k=2}^N b(\lambda, \nu_k) 
\otimes_{k=2}^N {}_k \langle -|.
\end{align}
From
\eqref{rel2} and \eqref{relation2}, one can derive
one recursive relation for the boundary correlation functions
between different lattice sizes \cite{CP2}
\begin{align}
\mathcal{F}_N^{(r, -, \epsilon_2, \cdots, \epsilon_{s})}
(\{ \lambda \}, \{ \nu \})=&\prod_{\beta=r+1}^N a(\lambda_\beta, \nu_1)
\sum_{\alpha=1}^r c
\prod_{ \substack{
\beta=1 \\
\beta \neq \alpha }}^r b(\lambda_\beta, \nu_1)
\prod_{ \substack{
\beta=1 \\
\beta \neq \alpha}}^{r} f(\lambda_\alpha, \lambda_\beta)
\prod_{k=2}^N a(\lambda_\alpha, \nu_k) \nonumber \\
&\times \mathcal{F}_{N-1}^{(r-1, \epsilon_2, \cdots, \epsilon_{s})}
(\{ \lambda \} \backslash \lambda_\alpha, \{ \nu \} \backslash \nu_1).
\label{recursive}
\end{align}
We can obtain another recursive relation
from \eqref{rel} and \eqref{relation3}
\begin{align}
\mathcal{F}_N^{(r, +, \epsilon_2, \cdots, \epsilon_{s})}
(\{ \lambda \}, \{ \nu \})
=&\prod_{\beta=1}^r b(\lambda_\beta, \nu_1)
\sum_{\alpha=r+1}^N c
\prod_{ \substack{
\beta=r+1 \\
\beta \neq \alpha }}^N a(\lambda_\beta, \nu_1)
\prod_{ \substack{
\beta=r+1 \\
\beta \neq \alpha}}^{N} f(\lambda_\beta, \lambda_\alpha)
\prod_{k=2}^N b(\lambda_\alpha, \nu_k) \nonumber \\
&\times \mathcal{F}_{N-1}^{(r, \epsilon_2, \cdots, \epsilon_{s})}
(\{ \lambda \} \backslash \lambda_\alpha, \{ \nu \} \backslash \nu_1).
\label{recursive2}
\end{align}
Solving these two recursive relations
\eqref{recursive} and \eqref{recursive2},
one obtains the general expression
for the boundary correlation functions as
\begin{align}
&\mathcal{F}_N^{(r, \epsilon_1, \cdots, \epsilon_{s})}
(\{ \lambda \}, \{ \nu \})
\nonumber \\
=&\frac{1}{\det M(\{ \lambda \}, \{ \nu \})}
\prod_{j=1}^s
\frac{\prod_{k=j+1}^N d(\nu_j, \nu_k)}
{\prod_{\beta=1}^r a(\lambda_\beta, \nu_j)
\prod_{\beta=r+1}^N b(\lambda_\beta, \nu_j)}
\sum_{\alpha_1 \in S_{\epsilon_1}^{N,r}}
\sum_{\substack{\alpha_2 \in S_{\epsilon_2}^{N,r} \\
\alpha_2 \neq \alpha_1}} \cdots
\sum_{\substack{\alpha_s \in S_{\epsilon_s}^{N,r} \\
\alpha_s \neq \alpha_1, \cdots, \alpha_{s-1}}}
\nonumber \\
&\times (-1)^{
\sum_{1 \le j < k \le s} \chi(\alpha_k, \alpha_j)
+\sum_{k=1}^s(\alpha_k-1-r(\epsilon_k+1)/2)
+\sum_{k=1}^s(\epsilon_k+1)(N-k)/2} \nonumber \\
&\times
\prod_{j=1}^s H_r^{\epsilon_j}(\lambda_{\alpha_j})
\prod_{1 \le j < k \le s}
E^{\epsilon_j \epsilon_k}(\lambda_{\alpha_j}, \lambda_{\alpha_k},
\nu_j, \nu_k)
\det M(\{ \lambda \} \backslash 
\{ \lambda_{\alpha_1}, \cdots, \lambda_{\alpha_s} \},
\{ \nu \} \backslash
\{ \nu_1, \cdots, \nu_s  \}
),
\label{boundarycorrelationfunction}
\end{align}
where
$S_-^{N,r}=\{1, \cdots, r \}, S_+^{N,r}=\{r+1, \cdots, N \}, 
e(\lambda, \nu)=\sh(\lambda-\nu+\eta)$,
\begin{align}
&H_r^-(\lambda)=\frac{\prod_{\beta=1}^r e(\lambda_\beta, \lambda)
\prod_{\beta=r+1}^N d(\lambda_\beta, \lambda)}
{\prod_{k=1}^N b(\lambda, \nu_k)},  \\
&H_r^+(\lambda)=\frac{\prod_{\beta=1}^r d(\lambda_\beta, \lambda)
\prod_{\beta=r+1}^N e(\lambda, \lambda_\beta)}
{\prod_{k=1}^N a(\lambda, \nu_k)},  \\
&E^{-+}(\lambda_{\alpha_j}, \lambda_{\alpha_k},
\nu_j, \nu_k)=\frac{a(\lambda_{\alpha_j}, \nu_k)a(\lambda_{\alpha_k}, \nu_j)}
{d(\lambda_{\alpha_j}, \lambda_{\alpha_k})},  \\
&E^{--}(\lambda_{\alpha_j}, \lambda_{\alpha_k},
\nu_j, \nu_k)=\frac{a(\lambda_{\alpha_j}, \nu_k)b(\lambda_{\alpha_k}, \nu_j)}
{e(\lambda_{\alpha_j}, \lambda_{\alpha_k})},  \\
&E^{++}(\lambda_{\alpha_j}, \lambda_{\alpha_k},
\nu_j, \nu_k)=\frac{b(\lambda_{\alpha_j}, \nu_k)a(\lambda_{\alpha_k}, \nu_j)}
{e(\lambda_{\alpha_k}, \lambda_{\alpha_j})},  \\
&E^{+-}(\lambda_{\alpha_j}, \lambda_{\alpha_k},
\nu_j, \nu_k)=\frac{b(\lambda_{\alpha_j}, \nu_k)b(\lambda_{\alpha_k}, \nu_j)}
{d(\lambda_{\alpha_j}, \lambda_{\alpha_k})},
\end{align}
and $\chi(\beta, \alpha)=1$ for $\beta > \alpha$ and 0 otherwise.
The proof is given in the Appendix.
As a special case ($\epsilon_j=-, j=1, \cdots, s$),
The boundary correlation function reduces to 
the emptiness formation probability \cite{CP2} (cf. \cite{KIB}),
which gives the probability of finding a sequence of
all spins down of length $s$ from the left boundary.

\section{Nineteen vertex model}
In this section, we consider the nineteen (spin-1 or
Fateev-Zamolodchikov) vertex model.
The nineteen vertex model can be constructed from the 
gauge transformed spin-1/2 $R$-matrix
\begin{align}
R_{jk}^+(\lambda, \nu)
=\phi_{j}(\lambda) \phi_{k}(\nu)
R_{jk}(\lambda, \nu)
\phi_{j}^{-1}(\lambda) \phi_{k}^{-1}(\nu),
\label{gaugespinonehalf}
\end{align}
where $\phi(\lambda)=\mathrm{diag}(1, \mathrm{e}^{\lambda})$
and the projection operator
\begin{align}
P=&\left(
\begin{array}{cccc}
1 & 0 & 0 & 0 \\
0 & \frac{ \mathrm{e}^\eta}{2 \ch \eta} & 
\frac{1}{2 \ch \eta} & 0 \\
0 & \frac{1}{2 \ch \eta} &
\frac{ \mathrm{e}^{-\eta}}{2 \ch \eta} & 0 \\
0 & 0 & 0 & 1
\end{array}
\right),
\end{align}
The basis (dual basis) of the spin-1 representation
$|1 \rangle, |0 \rangle, |-1 \rangle$ 
($\langle 1|, \langle 0|, \langle -1|$) is 
given in terms of basis of spin-1/2 representation as
$|1 \rangle=|+ \rangle \otimes |+ \rangle,
|0 \rangle=(1+\mathrm{e}^{-2 \eta})^{-1/2}
(|+ \rangle \otimes |- \rangle + 
\mathrm{e}^{-\eta} |- \rangle \otimes |+ \rangle),
|-1 \rangle=|- \rangle \otimes |- \rangle
$.
The gauge transformed spin-1 $R$-matrix
can be constructed as
\cite{KRS,DWA,DJKMO}
\begin{align}
R_{JK}^{1+}(z,w)=&P_{2K-1,2K}P_{2J,2J-1}
R_{2J,2K}^+(z+\eta,w) R_{2J,2K-1}^+(z+\eta, w+\eta) \nonumber \\
&\times R_{2J-1,2K}^+(z,w)R_{2J-1,2K-1}^+(z,w+\eta)
P_{2J,2J-1}P_{2K-1,2K}.
\end{align}
The symmetric spin-1 $R$-matrix can be obtained from
$R_{12}^{1+}(z,w)$ by gauging out factors as
\begin{align}
R_{JK}^{1}(z, w)
=\Phi_{J}^{-1}(z) \Phi_{K}^{-1}(w)
R_{JK}^{1+}(z, w)
\Phi_{J}(z) \Phi_{K}(w),
\label{gaugespinone}
\end{align}
where $\Phi(z)=\mathrm{diag}(1, \mathrm{e}^{z}, \mathrm{e}^{2z})$.

For the nineteen vertex model
on a $N \times N$ lattice with domain wall boundary condition,
all spins are aligned +1 at the bottom and right boundaries,
and $-1$ at the top and left boundaries.
At the intersection of the $\alpha$-th row (from the bottom)
and the $k$-th column (from the left), the statistical weight
$L_{\alpha k}^1(z_\alpha,
w_k)=
R_{\alpha k}^1(z_{\alpha}-\eta/2, w_k)
$
is associated.
We also set $L_{\alpha k}^{1+}(z_\alpha,
w_k)=
R_{\alpha k}^{1+}(z_{\alpha}-\eta/2, w_k)
$,
$L_{\alpha k}^{1/2}(z_\alpha,
w_k)=
R_{\alpha k}(z_{\alpha}-\eta/2, w_k)
$,
$L_{\alpha k}^{1/2+}(z_\alpha,
w_k)=
R_{\alpha k}^{+}(z_{\alpha}-\eta/2, w_k)
$
 for later convenience.

We consider the boundary correlation functions for
this nineteen vertex model
\begin{align}
F_N^{1(r, \delta_1, \cdots, \delta_{s})}
(\{ z \}, \{ w \})
&=\frac{
\widetilde{F}_N^{1(r, \delta_1, \cdots, \delta_{s})}
(\{ z \}, \{ w \})
}
{Z_{N}^1(\{ z \}, \{ w \})},
\label{spinoneboundarycorrelation}
\\
Z_N^1(\{ z \}, \{ w \})
&= {}_a\langle 1|| {}_q\langle -1|| 
\prod_{\alpha,k=1}^N L^1_{\alpha k}
(z_{\alpha}, w_k)||-1 \rangle_a ||1 \rangle_q,
\label{spinonepartitionfunction}
\\
\widetilde{F}_N^{1(r, \delta_1, \cdots, \delta_{s})}
(\{ z \}, \{ w \})
&= {}_a\langle 1|| {}_q\langle -1|| 
\prod_{\alpha=r+1}^N \prod_{k=1}^N L^1_{\alpha k}
(z_{\alpha}, w_k)
\prod_{k=1}^s \pi_k^{\delta_k}
\prod_{\alpha=1}^r \prod_{k=1}^N L^1_{\alpha k}
(z_{\alpha}, w_k)
||-1 \rangle_a ||1 \rangle_q,
\label{spinonenumerator}
\end{align}
where $\pi_k^{\delta_k}=|\delta_k \rangle_k {}_k \langle \delta_k|
, \delta_k=1,0,-1$ and
$ ||1 \rangle=\otimes_{k=1}^N | 1 \rangle_k ,
  ||-1 \rangle=\otimes_{k=1}^N | -1 \rangle_k,
\langle 1 ||=\otimes_{k=1}^N  {}_k \langle 1|,
\langle -1 ||=\otimes_{k=1}^N  {}_k \langle -1|$.

We show that the above boundary correlation functions can be reduced
to those for the six vertex model calculated in the previous section.
Instead of directly dealing with \eqref{spinoneboundarycorrelation},
we consider
\begin{align}
F_N^{1+(r, \delta_1, \cdots, \delta_{s})}
(\{ z \}, \{ w \})
&=\frac{
\widetilde{F}_N^{1+(r, \delta_1, \cdots, \delta_{s})}
(\{ z \}, \{ w \})
}
{Z_{N}^{1+}(\{ z \}, \{ w \})}, \\
Z_{N}^{1+}(\{ z \}, \{ w \})&=Z_{N}^1(\{ z \}, \{ w \})|_{L^1
\to L^{1+}}, \\
\widetilde{F}_N^{1+(r, \delta_1, \cdots, \delta_{s})}
(\{ z \}, \{ w \})&=\widetilde{F}_N^{1(r, \delta_1, \cdots, \delta_{s})}
(\{ z \}, \{ w \})|_{L^1
 \to L^{1+}}.
\end{align}
We also define $F_N^{1/2(r, \epsilon_1, \cdots, \epsilon_{s})}$
and $F_N^{1/2+(r, \epsilon_1, \cdots, \epsilon_{s})}$ as well,
replacing $L^1$ by $L^{1/2}$ and $L^{1/2+}$, respectively. \\
From
\eqref{gaugespinone}, one can see
\begin{align}
F_N^{1+(r, \delta_1, \cdots, \delta_{s})}
(\{ z \}, \{ w \})
=F_N^{1(r, \delta_1, \cdots, \delta_{s})}
(\{ z \}, \{ w \}),
\label{gaugecorrelation}
\end{align}
since
\begin{align}
Z_{N}^{1+}(\{ z \}, \{ w \})&=
\mathrm{e}^{2 \sum_{k=1}^N w_k -2 \sum_{\alpha=1}^N(z_\alpha-\eta/2)}
Z_{N}^1(\{ z \}, \{ w \}), \\
\widetilde{F}_N^{1+(r, \delta_1, \cdots, \delta_{s})}
(\{ z \}, \{ w \})&=
\mathrm{e}^{2 \sum_{k=1}^N w_k -2 \sum_{\alpha=1}^N(z_\alpha-\eta/2)}
\widetilde{F}_N^{1(r, \delta_1, \cdots, \delta_{s})}
(\{ z \}, \{ w \}).
\end{align}
Thus, we can consider 
$F_N^{1+(r, \delta_1, \cdots, \delta_{s})}
(\{ z \}, \{ w \})
$ instead, which is easier to handle than 
$F_N^{1(r, \delta_1, \cdots, \delta_{s})}
(\{ z \}, \{ w \})$ itself.

We reduce $F_N^{1+(r, \delta_1, \cdots, \delta_{s})}
(\{ z \}, \{ w \})
$ to the boundary correlation functions of 
the six vertex model by use of fusion.
The following relations are used.
\begin{align}
&P_{2K-1,2K}
R_{2J,2K}^+(z+\eta,w) R_{2J,2K-1}^+(z+\eta, w+\eta)
R_{2J-1,2K}^+(z,w)R_{2J-1,2K-1}^+(z,w+\eta)
P_{2K-1,2K} \nonumber \\ 
&=P_{2K-1,2K}
R_{2J,2K}^+(z+\eta,w) R_{2J,2K-1}^+(z+\eta, w+\eta)
R_{2J-1,2K}^+(z,w)R_{2J-1,2K-1}^+(z,w+\eta), \label{fusion1} \\
&P_{2J,2J-1}
R_{2J,2K}^+(z+\eta,w) R_{2J,2K-1}^+(z+\eta, w+\eta)
R_{2J-1,2K}^+(z,w)R_{2J-1,2K-1}^+(z,w+\eta)
P_{2J,2J-1} \nonumber \\
&=P_{2J,2J-1}
R_{2J,2K}^+(z+\eta,w) R_{2J,2K-1}^+(z+\eta, w+\eta)
R_{2J-1,2K}^+(z,w)R_{2J-1,2K-1}^+(z,w+\eta), \label{fusion2} \\
&P^2=P, \
P|\pm 1 \rangle=|\pm \rangle \otimes |\pm \rangle, \
\langle \pm 1|P=\langle \pm | \otimes \langle \pm|, \label{fusion3} \\
&\pi_k^{\delta_k} P_{2k-1,2k}
=P_{2k-1,2k} \sum_{\epsilon_{2k-1}, \epsilon_{2k}=\pm}
C_{\epsilon_{2k-1} \epsilon_{2k}}^{\delta_k}
\pi_{2k-1}^{\epsilon_{2k-1}} \pi_{2k}^{\epsilon_{2k}}, \label{fusion4}
\end{align}
where $C_{++}^1=C_{--}^{-1}=C_{+-}^0=C_{-+}^0=1$ and 0 otherwise.

First, utilizing \eqref{fusion1}, \eqref{fusion2} and \eqref{fusion3},
one has
\begin{align}
\widetilde{F}_N^{1+(r, \delta_1, \cdots, \delta_{s})}
(\{ z \}, \{ w \})
=&{}_a \langle 1|| {}_q \langle -1|| 
\prod_{k=1}^N P_{2k-1, 2k}
\prod_{\alpha=r+1}^N \widetilde{T}_\alpha(z_\alpha, \{ w \})
\prod_{k=1}^s P_{2k-1, 2k} \pi_k^{\delta_k} P_{2k-1, 2k}
\nonumber \\
&\times \prod_{\alpha=1}^r \widetilde{T}_\alpha(z_\alpha, \{ w \})
||- \rangle_a ||+ \rangle_q,
\end{align}
where
$ ||+ \rangle=\otimes_{k=1}^{2N} | + \rangle_k ,
  ||- \rangle=\otimes_{k=1}^{2N} | - \rangle_k,
\langle + ||=\otimes_{k=1}^{2N}  {}_k \langle +|,
\langle - ||=\otimes_{k=1}^{2N}  {}_k \langle -|$ and
\begin{align}
\widehat{T}_\alpha(z_\alpha, \{ w \})=&
\prod_{k=1}^N L^{1/2+}_{2 \alpha, 2k}(z_\alpha+\eta, w_k)
L^{1/2+}_{2 \alpha, 2k-1}(z_\alpha+\eta, w_k+\eta) \nonumber \\
&\times 
\prod_{k=1}^N L^{1/2+}_{2 \alpha-1, 2k}(z_\alpha, w_k)
L^{1/2+}_{2 \alpha-1, 2k-1}(z_\alpha, w_k+\eta),
\\
\widetilde{T}_\alpha(z_\alpha, \{ w \})=&
P_{2 \alpha, 2 \alpha-1} \widehat{T}_\alpha(z_\alpha, \{ w \}).
\end{align}
Applying \eqref{fusion4},
we have
\begin{align}
\widetilde{F}_N^{1+(r, \delta_1, \cdots, \delta_{s})}
(\{ z \}, \{ w \})
=&{}_a \langle 1|| {}_q \langle -1|| 
\prod_{k=1}^N P_{2k-1, 2k}
\prod_{\alpha=r+1}^N \widetilde{T}_\alpha(z_\alpha, \{ w \})
\nonumber \\
&\times \prod_{k=1}^s \{ P_{2k-1, 2k} 
\sum_{\epsilon_{2k-1}, \epsilon_{2k}=\pm}
C_{\epsilon_{2k-1} \epsilon_{2k}}^{\delta_k}
\pi_{2k-1}^{\epsilon_{2k-1}} \pi_{2k}^{\epsilon_{2k}}
\}
\nonumber \\
&\times \prod_{\alpha=1}^r \widetilde{T}_\alpha(z_\alpha, \{ w \})
||- \rangle_a ||+ \rangle_q.
\end{align}
Using \eqref{fusion1} and \eqref{fusion3},
one gets
\begin{align}
\widetilde{F}_N^{1+(r, \delta_1, \cdots, \delta_{s})}
(\{ z \}, \{ w \})
=&\sum_{\epsilon_{1}, \cdots, \epsilon_{2s}=\pm}
\prod_{k=1}^s C_{\epsilon_{2k-1} \epsilon_{2k}}^{\delta_k}
{}_a \langle +|| {}_q \langle -|| 
\prod_{\alpha=r+1}^N \widehat{T}_\alpha(z_\alpha, \{ w \})
\nonumber \\
&\times
\prod_{k=1}^s \pi_{2k-1}^{\epsilon_{2k-1}} \pi_{2k}^{\epsilon_{2k}}
\prod_{\alpha=1}^r \widehat{T}_\alpha(z_\alpha, \{ w \})
||- \rangle_a ||+ \rangle_q \nonumber \\
=&\sum_{\epsilon_{1}, \cdots , \epsilon_{2s}=\pm}
\prod_{k=1}^s
C_{\epsilon_{2k-1} \epsilon_{2k}}^{\delta_k}
\widetilde{F}_{2N}^{1/2+(2r, \epsilon_1, \cdots, \epsilon_{2s})}
(\{ \bar{z} \}, \{ \bar{w} \}),
\end{align}
where
$\{ \bar{z} \}=\{ z_1, z_1+\eta, z_2, z_2+\eta, \cdots, z_N, z_N+\eta \}$,
$\{ \bar{w} \}=\{ w_1+\eta, w_1, w_2+\eta, w_2, \cdots, w_N+\eta, w_N \}$.

As the simplest case, one has \cite{CFK}
\begin{align}
Z_N^{1+}
(\{ z \}, \{ w \})
=Z_{2N}^{1/2+}
(\{ \bar{z} \}, \{ \bar{w} \}).
\end{align}
Thus we have
\begin{align}
F_N^{1+(r, \delta_1, \cdots, \delta_{s})}
(\{ z \}, \{ w \})
=&\sum_{\epsilon_{1}, \cdots, \epsilon_{2s}=\pm}
\prod_{k=1}^s C_{\epsilon_{2k-1} \epsilon_{2k}}^{\delta_k}
F_{2N}^{1/2+(2r, \epsilon_1, \cdots, \epsilon_{2s})}
(\{ \bar{z} \}, \{ \bar{w} \}).
\label{reduction}
\end{align}
From
\eqref{gaugespinonehalf}, one can see
\begin{align}
F_{2N}^{1/2+(2r, \epsilon_1, \cdots, \epsilon_{2s})}
(\{ \bar{z} \}, \{ \bar{w} \})
=F_N^{1/2(2r, \epsilon_1, \cdots, \epsilon_{2s})}
(\{ \bar{z} \}, \{ \bar{w} \}), \label{gaugecorrelation2}
\end{align}
since
\begin{align}
Z_{2N}^{1/2+}(\{ \bar{z} \}, \{ \bar{w} \})&=
\mathrm{e}^{2 \sum_{k=1}^N w_k -2 \sum_{\alpha=1}^N z_\alpha+N \eta}
Z_{2N}^{1/2}(\{ \bar{z} \}, \{ \bar{w} \}), \\
\widetilde{F}_{2N}^{1/2+(2r, \epsilon_1, \cdots, \epsilon_{2s})}
(\{ \bar{z} \}, \{ \bar{w} \})&=
\mathrm{e}^{2 \sum_{k=1}^N w_k -2 \sum_{\alpha=1}^N z_\alpha+N \eta}
\widetilde{F}_{2N}^{1/2(2r, \epsilon_1, \cdots, \epsilon_{2s})}
(\{ \bar{z} \}, \{ \bar{w} \}).
\end{align}
Combining \eqref{gaugecorrelation}, \eqref{reduction} and 
\eqref{gaugecorrelation2}, one finally has
\begin{align}
F_N^{1(r, \delta_1, \cdots, \delta_{s})}
(\{ z \}, \{ w \})
=&\sum_{\epsilon_{1}, \cdots, \epsilon_{2s}=\pm}
\prod_{k=1}^s C_{\epsilon_{2k-1} \epsilon_{2k}}^{\delta_k}
F_{2N}^{1/2(2r, \epsilon_1, \cdots, \epsilon_{2s})}
(\{ \bar{z} \}, \{ \bar{w} \}),
\label{reduction2}
\end{align}
which means that the boundary correlation functions for the
nineteen vertex model on an $N \times N$ lattice
with spectral parameters $\{ z \}, \{ w \}$
can be reduced to those for the six
vertex model on a $2N \times 2N$ lattice
with spectral parmeters $\{ \bar{z} \}, \{ \bar{w} \}$
. Note that
$F_{2N}^{1/2(2r, \epsilon_1, \cdots, \epsilon_{2s})}
(\{ \bar{z} \}, \{ \bar{w} \})$ is exactly
$\mathcal{F}_{2N}^{(2r, \epsilon_1, \cdots, \epsilon_{2s})}
(\{ \bar{z} \}, \{ \bar{w} \})$
in the previous section since the corresponding Boltzmann weights are
different just by an overall factor,
which do not affect correlation functions.
\section{Homogeneous limit of the emptiness formation probability}
Let us consider the homogeneous limit of 
the emptiness formation probability (EFP) for the
nineteen vertex model.
As a special case of \eqref{reduction2}, one has
\begin{align}
F_N^{1(r, (-)^s)}
(\{ z \}, \{ w \})
=
\mathcal{F}_{2N}^{(2r, (-)^{2s})}
(\{ \bar{z} \}, \{ \bar{w} \}),
\end{align}
i.e., the EFP of length $s$
for the nineteen vertex model
with spectral parameters
$\{ z \}, \{ w \}$
reduces to the EFP of length $2s$ for the six vertex model
with spectral parameters
$\{ \bar{z} \}=\{ z_1, z_1+\eta, z_2, z_2+\eta, \cdots, z_N, z_N+\eta \}$,
$\{ \bar{w} \}=\{ w_1+\eta, w_1, w_2+\eta, w_2, \cdots, w_N+\eta, w_N \}$.
Let us set $z_j$ as $z_j=z+\xi_j$.
One finds
that $\mathcal{F}_{2N}^{(2r, (-)^{2s})}
(\{ \bar{z} \}, \{ \bar{w} \})$
can be expressed in the determinant form as
\begin{align}
\mathcal{F}_{2N}^{(2r, (-)^{2s})}
(\{ \bar{z} \}, \{ \bar{w} \})
=\frac{X_1}{\det M(\{ \bar{z} \}, \{ \bar{w} \})}
\det \Psi \frac{X_2 X_3}{\prod_{1 \le j < k < 2s}e(z+\epsilon_j, z+\epsilon_k)}
\Big|_{\epsilon_1= \cdots = \epsilon_{2s}=0},
\end{align}
where
\begin{align}
X_1=&
\frac{ \prod_{j=1}^s {[}
\prod_{k=j}^N d(w_j+\eta, w_k)
\prod_{k=j+1}^N d^2(w_j, w_k)
\prod_{k=j+1}^N d(w_j, w_k+\eta)
{]}
}
{
\prod_{j=1}^s {[}
\prod_{\beta=1}^r
a(z_\beta+\eta, w_j)a^2(z_\beta, w_j)a(z_\beta,w_j+\eta)
{]}
{[}
\prod_{\beta=r+1}^N
b(z_\beta+\eta, w_j)b^2(z_\beta, w_j)b(z_\beta,w_j+\eta)
{]}
}, \\
X_2=&
\frac{
\prod_{j=1}^{2s}
{[}
\prod_{\beta=1}^r e(z_\beta, z+\epsilon_j)
e(z_\beta+\eta, z+\epsilon_j)
{]}
{[}
\prod_{\beta=r+1}^N d(z_\beta, z+\epsilon_j)
d(z_\beta+\eta, z+\epsilon_j)
{]}
}
{
\prod_{j=1}^{2s}
{[} \prod_{k=1}^N
b(z+\epsilon_j, w_k)b(z+\epsilon_j, w_k+\eta)
{]}
}, \\
X_3=&\prod_{j=1}^s {[}
\prod_{k=j}^s a(z+\epsilon_{2j-1}, w_k)
\prod_{k=j+1}^s a(z+\epsilon_{2j-1}, w_k+\eta)
 {]} \nonumber \\
&\times
\prod_{j=1}^{s-1} {[}
\prod_{k=j+1}^s a(z+\epsilon_{2j}, w_k)
\prod_{k=j+1}^s a(z+\epsilon_{2j}, w_k+\eta)
 {]} \nonumber \\
&\times \prod_{k=2}^s {[}
\prod_{j=1}^{k-1} b(z+\epsilon_{2k-1}, w_j)
\prod_{j=1}^{k-1} b(z+\epsilon_{2k-1}, w_j+\eta)
 {]} \nonumber \\
&\times
\prod_{k=1}^{s} {[}
\prod_{j=1}^{k-1} b(z+\epsilon_{2k}, w_j)
\prod_{j=1}^k b(z+\epsilon_{2k}, w_j+\eta)
 {]},
\end{align}
and $\Psi$ is a $2N \times 2N$ matrix whose $(j,k)$-th
block matrix element is given by
\begin{align}
\left(
\begin{array}{cc}
\mathrm{exp}(\xi_j \partial_{\epsilon_{2k-1}}) &
\mathrm{exp}(\xi_j \partial_{\epsilon_{2k}})  \\
\mathrm{exp}((\xi_j+\eta) \partial_{\epsilon_{2k-1}}) &
\mathrm{exp}((\xi_j+\eta) \partial_{\epsilon_{2k}})  
\end{array}
\right),
\end{align}
for $j=1, \cdots, N, k=1, \cdots, s$
and
\begin{align}
\left(
\begin{array}{cc}
\varphi(z_j, w_k+\eta) &
\varphi(z_j, w_k)  \\
\varphi(z_j+\eta, w_k+\eta) &
\varphi(z_j+\eta, w_k)  
\end{array}
\right),
\end{align}
for $j=1, \cdots, N, k=s+1, \cdots, N$.

Now let us take the homogeneous limit
by putting $\xi_j, w_j , j=1, \cdots, N$ to zero in the order 
$w_1 \to 0, \dots, w_N \to 0, \xi_1 \to 0, \dots, \xi_N \to 0$.
We have
\begin{align}
F_{N}^{1(r,(-)^s)}
=\mathcal{F}_{2N}^{(2r, (-)^{2s})}
=\frac{Y_1}{\det m}
\det \psi \frac{Y_2 Y_3}{\prod_{1 \le j < k < 2s}
\sh(\epsilon_j-\epsilon_k+\eta)}
\Big|_{\epsilon_1= \cdots = \epsilon_{2s}=0},
\end{align}
where
\begin{align}
Y_1=&\frac{(-1)^{sN-s(s+1)/2}
\{ \prod_{j=1}^s (N-j)! \}^2
d^{2sN-s^2}(\eta,0)
}
{
a^{rs}(z+\eta,0)b^{(N-r)s}(z+\eta,0)
a^{2rs}(z,0)b^{2(N-r)s}(z,0)
a^{rs}(z-\eta,0)b^{(N-r)s}(z-\eta,0)
}, \\
Y_2=&\prod_{j=1}^{2s}
\frac{\sh^r(-\epsilon_j+2 \eta) \sh^N(-\epsilon_j+\eta)
\sh^{N-r}(-\epsilon_j)}{\sh^N(\epsilon_j+z-\eta/2)
\sh^N(\epsilon_j+z-3 \eta/2)}, \\
Y_3=&\prod_{j=1}^s \sh^{s-j+1}(z+\epsilon_{2j-1}+\eta/2)
\sh^{s-j}(z+\epsilon_{2j-1}-\eta/2) \nonumber \\
&\times
\prod_{j=1}^{s-1} \sh^{s-j}(z+\epsilon_{2j}+\eta/2)
\sh^{s-j}(z+\epsilon_{2j}-\eta/2) \nonumber \\
&\times
\prod_{k=2}^s \sh^{k-1}(z+\epsilon_{2k-1}-\eta/2)
\sh^{k-1}(z+\epsilon_{2k-1}-3 \eta/2) \nonumber \\
&\times
\prod_{k=1}^{s} \sh^{k-1}(z+\epsilon_{2k}-\eta/2)
\sh^{k}(z+\epsilon_{2k}-3 \eta/2),
\end{align}
$m$ is a $2N \times 2N$ matrix whose $(j,k)$-th
block matrix element is given by
\begin{align}
\left(
\begin{array}{cc}
\partial_z^{j+k-2} \varphi(z-\eta, 0) &
\partial_z^{j+k-2} \varphi(z, 0)  \\
\partial_z^{j+k-2} \varphi(z, 0) &
\partial_z^{j+k-2} \varphi(z+\eta, 0)  
\end{array}
\right),
\end{align}
for $j,k=1, \cdots, N$,
and $\psi$ is a $2N \times 2N$ matrix whose $(j,k)$-th
block matrix element is given by
\begin{align}
\left(
\begin{array}{cc}
\partial_{\epsilon_{2k-1}}^{j+k-2} &
\partial_{\epsilon_{2k}}^{j+k-2}  \\
\mathrm{exp}(\eta \partial_{\epsilon_{2k-1}}
) \partial_{\epsilon_{2k-1}}^{j+k-2} &
\mathrm{exp}(\eta \partial_{\epsilon_{2k}}
) \partial_{\epsilon_{2k}}^{j+k-2}
\end{array}
\right),
\end{align}
for $j=1, \cdots, N, k=1, \cdots, s$
and
\begin{align}
\left(
\begin{array}{cc}
\partial_z^{j+k-s-2} \varphi(z-\eta, 0) &
\partial_z^{j+k-s-2} \varphi(z, 0)  \\
\partial_z^{j+k-s-2} \varphi(z, 0) &
\partial_z^{j+k-s-2} \varphi(z+\eta, 0)  
\end{array}
\right),
\end{align}
for $j=1, \cdots, N, k=s+1, \cdots, N$.

\section{Conclusion}
In this paper, we considered correlation functions
for the six and nineteen vertex models on an $N \times N$ lattice
with domain wall boundary conditions. 
For the six vertex model, we derived the general expression for the
boundary correlation function by solving two recursive relations
obtained by the quantum inverse scattering method.
The result includes the boundary one point functions and 
emptiness formation probability as special cases.

For the nineteen vertex (Fateev-Zamolodchikov) model, by use of fusion,
we have shown that the boundary correlation functions reduce to
those for the six vertex model. In particular,
the emptiness formation probability of length $s$
for the nineteen vertex model on an $N \times N$ lattice
reduces to that of length $2s$ for the six vertex model
on a $2N \times 2N$ lattice with appropriate spectral parameters.

The correlation functions "off the boundary"
can in principle be expressed as a linear sum of 
the boundary correlation functions obtained in this paper.
However, since this means we need to sum over
intermediate spin states, the expression gets complicated.
Simplifying the expression of the correlation functions
"off the boundary" is an important problem to be considered in
the future.

Another interesting problem is to extend the analysis to
other models or boundary conditions such as higher rank models,
Felderhof model, reflecting end, etc.

\section*{Appendix}
\renewcommand{\theequation}{A.\arabic{equation}}
\setcounter{equation}{0}
We prove \eqref{boundarycorrelationfunction} by induction.
We show the expression holds for lattice size $N$ and
$\epsilon_1=-$ from the recursive relation \eqref{recursive}.
One can similarly show for $\epsilon_1=+$ from \eqref{recursive2}.
Suppose \eqref{boundarycorrelationfunction} holds for lattice size $N-1$.
For $\alpha_1 \in S_{-}^{N,r}$, one has
\begin{align}
&\mathcal{F}_{N-1}^{(r-1, \epsilon_2, \cdots, \epsilon_{s})}
(\{ \lambda \} \backslash \lambda_{\alpha_1}, 
\{ \nu \} \backslash \nu_1)
\nonumber \\
=&\frac{1}{\det M(\{ \lambda \} \backslash \lambda_{\alpha_1},
\{ \nu \} \backslash \nu_1)}
\prod_{j=2}^s
\frac{\prod_{k=j+1}^N d(\nu_j, \nu_k)}
{\prod_{\substack{ \beta=1 \\ \beta \neq \alpha_1 }}^r
a(\lambda_\beta, \nu_j)
\prod_{\beta=r+1}^N b(\lambda_\beta, \nu_j)}
\sum_{\substack{\alpha_2 \in S_{\epsilon_2}^{N,r} \\
\alpha_2 \neq \alpha_1}} \cdots
\sum_{\substack{\alpha_s \in S_{\epsilon_s}^{N,r} \\
\alpha_s \neq \alpha_1, \cdots, \alpha_{s-1}}}
\nonumber \\
&\times (-1)^{
\sum_{1 \le j < k \le s} \chi(\alpha_k, \alpha_j)
+\sum_{k=2}^s(\alpha_k-1-r(\epsilon_k+1)/2)
+\sum_{k=2}^s(\epsilon_k+1)(N-k)/2} \nonumber \\
&\times
\prod_{j=2}^s H_r^{\epsilon_j}(\lambda_{\alpha_j})
m^{\epsilon_j}(\lambda_{\alpha_1}, \lambda_{\alpha_j}, \nu_1)
\prod_{2 \le j < k \le s}
E^{\epsilon_j \epsilon_k}(\lambda_{\alpha_j}, \lambda_{\alpha_k},
\nu_j, \nu_k) \nonumber \\
&\times \det M(\{ \lambda \} \backslash 
\{ \lambda_{\alpha_1}, \cdots, \lambda_{\alpha_s} \},
\{ \nu \} \backslash
\{ \nu_1, \cdots, \nu_s  \}
), \label{sizeminusone}
\end{align}
where
\begin{align}
m^{+}(\lambda_{\alpha_1}, \lambda_{\alpha_j}, \nu_1)
=\frac{a(\lambda_{\alpha_j}, \nu_1)}
{d(\lambda_{\alpha_1}, \lambda_{\alpha_j})}, \
m^{-}(\lambda_{\alpha_1}, \lambda_{\alpha_j}, \nu_1)
=\frac{b(\lambda_{\alpha_j}, \nu_1)}
{e(\lambda_{\alpha_1}, \lambda_{\alpha_j})}.
\end{align}
We also have the follwing recursive relation \cite{Izergin,ICK,CP2} for the
parititon function
\begin{align}
\frac{\mathcal{Z}_{N-1}(\{ \lambda \} \backslash \lambda_\alpha, 
\{ \nu \} \backslash \nu_1  )}
{\mathcal{Z}_N(\{ \lambda \}, \{ \nu \})}
=&\frac{(-1)^{\alpha-1}}{a(\lambda_\alpha, \nu_1)b(\lambda_\alpha, \nu_1)}
\prod_{\substack{ \beta=1 \\ \beta \neq \alpha }}^N 
\frac{d(\lambda_\beta, \lambda_\alpha)}{a(\lambda_\beta, \nu_1)b(\lambda_\beta, \nu_1)} \prod_{k=2}^N \frac{d(\nu_1, \nu_k)}{a(\lambda_\alpha, \nu_k)
b(\lambda_\alpha, \nu_k)}
\nonumber \\
&\times \frac{\mathrm{det}M(\{ \lambda \} \backslash \lambda_\alpha, 
\{ \nu \} \backslash \nu_1  )}
{\mathrm{det}M(\{ \lambda \}, \{ \nu \})}.
\label{partitionrecursive}
\end{align}
Combining \eqref{recursive}, \eqref{sizeminusone} 
and \eqref{partitionrecursive}, one has
\begin{align}
&\mathcal{F}_{N}^{(r, -, \epsilon_2, \cdots, \epsilon_{s})}
(\{ \lambda \}, \{ \nu \})
\nonumber \\
=&\frac{1}{\det M(\{ \lambda \}, \{ \nu \})}
\prod_{j=1}^s
\frac{\prod_{k=j+1}^N d(\nu_j, \nu_k)}
{\prod_{\beta=1}^r
a(\lambda_\beta, \nu_j)
\prod_{\beta=r+1}^N b(\lambda_\beta, \nu_j)}
\sum_{\alpha_1 \in S_{-}^{N,r}}
\sum_{\substack{\alpha_2 \in S_{\epsilon_2}^{N,r} \\
\alpha_2 \neq \alpha_1}} \cdots
\sum_{\substack{\alpha_s \in S_{\epsilon_s}^{N,r} \\
\alpha_s \neq \alpha_1, \cdots, \alpha_{s-1}}}
\nonumber \\
&\times (-1)^{
\sum_{1 \le j < k \le s} \chi(\alpha_k, \alpha_j)
+\sum_{k=1}^s(\alpha_k-1-r(\epsilon_k+1)/2)
+\sum_{k=1}^s(\epsilon_k+1)(N-k)/2} \nonumber \\
&\times
\prod_{j=2}^s H_r^{\epsilon_j}(\lambda_{\alpha_j})
\prod_{2 \le j < k \le s}
E^{\epsilon_j \epsilon_k}(\lambda_{\alpha_j}, \lambda_{\alpha_k},
\nu_j, \nu_k) \det M(\{ \lambda \} \backslash 
\{ \lambda_{\alpha_1}, \cdots, \lambda_{\alpha_s} \},
\{ \nu \} \backslash
\{ \nu_1, \cdots, \nu_s  \}
) \nonumber \\
&\times 
\frac{c
\prod_{\substack{\beta=1 \\
\beta \neq \alpha_1}}^r f(\lambda_{\alpha_1}, \lambda_\beta)
\prod_{\substack{\beta=1 \\
\beta \neq \alpha_1}}^N d(\lambda_\beta, \lambda_{\alpha_1})
}{\prod_{k=1}^N b(\lambda_{\alpha_1}, \nu_k)}
\prod_{j=2}^s a(\lambda_{\alpha_1}, \nu_j)
m^{\epsilon_j}(\lambda_{\alpha_1}, \lambda_{\alpha_j}, \nu_1).
\label{almostexpression}
\end{align}
Since
\begin{align}
\frac{c
\prod_{\substack{\beta=1 \\
\beta \neq \alpha_1}}^r f(\lambda_{\alpha_1}, \lambda_\beta)
\prod_{\substack{\beta=1 \\
\beta \neq \alpha_1}}^N d(\lambda_\beta, \lambda_{\alpha_1})
}{\prod_{k=1}^N b(\lambda_{\alpha_1}, \nu_k)}
&=H_r^-(\lambda_{\alpha_1}), \nonumber \\
a(\lambda_{\alpha_1}, \nu_j)
m^{+}(\lambda_{\alpha_1}, \lambda_{\alpha_j}, \nu_1)
&=E^{-+}(\lambda_{\alpha_1}, \lambda_{\alpha_j},
\nu_1, \nu_j),
\nonumber \\
a(\lambda_{\alpha_1}, \nu_j)
m^{-}(\lambda_{\alpha_1}, \lambda_{\alpha_j}, \nu_1)
&=E^{--}(\lambda_{\alpha_1}, \lambda_{\alpha_j},
\nu_1, \nu_j),
\end{align}
one can see that
\eqref{almostexpression} is exactly the expression
\eqref{boundarycorrelationfunction} for $\epsilon_1=-$.

\end{document}